\documentstyle[aps,epsf]{revtex}
\begin{document}

\twocolumn

\title{Comment on: "Auger decay, Spin-exchange, and their
connection to Bose-Einstein condensation 
of excitons in $Cu_2O$ " \\
by 
G.M. Kavoulakis, A. Mysyrowicz ( cond-mat/0001438 )}

\author{A. B. Kuklov}
\address{ Dept. of Eng. and Physics,
College of  Staten Island, CUNY
     }

\maketitle

\vskip0.5 cm

In the very recent work \cite{KAV} a new mechanism 
of the interconversion of the triplet excitons
into singlet excitons in $Cu_2O$ has been suggested. In
accordance with it, two triplet excitons
with the opposite (internal)
 angular momenta may collide and interconvert into a pair 
of the singlet excitons. 
 Estimates presented in
\cite{KAV} show that such an interconversion
is the most effective channel for the
decrease of the triplet exciton density. This
questions the commonly accepted view 
that the Auger decay 
is the primary channel
for the decay (see in \cite{OHARA}).  
Furthermore, it has been pointed out in \cite{KAV}
that 
the actual rate of the Auger decay must be
several orders of magnitude less than 
it was previously calculated. 

In this comment, it is suggested that 
the mechanism \cite{KAV} leads to a verifiable
prediction about the rate of the decrease of the
triplet exciton density as function of the
polarization of the incident laser light
inducing the two-photon creation of the triplet excitons.
As a matter of fact, this rate can be greatly reduced
if the polarization of the laser light is properly
chosen.

First, it is worth noting that
a necessary condition 
for the triplet-singlet
interconversion \cite{KAV} is that
before the collision a pair of the triplet excitons
has total angular momentum $J=0$, that is, they form
a singlet state (invariant under the point
group rotations). Thus, the rate of the 
interconversion should be extremely sensitive
to the initial state of the excitonic
ensemble. If this state is a thermal mixture
with random orientation of the excitonic
spins, then each exciton can easily
find another one with the opposite $J_z$,
so that the collision between them would result
in a pair of the singlet excitons \cite{KAV}.
On the contrary, if initially excitonic spins 
were aligned, the collision induced
interconversion will be completely suppressed
because of the conservation of the angular
momentum of the colliding pairs. Thus,
aligning angular momenta  of the triplet 
excitons in one way or another should
prevent the triplet excitons from  
transforming into the singlet excitons
in accordance with the mechanism \cite{KAV}.
This property can be used as a test for
the mechanism \cite{KAV}.

One way for preparing
triplet excitons with preferential orientation
of their spins is a direct creation
of the coherent triplet excitons employed in
\cite{GOTO}. In this work, the triplet
excitons have been created by the two-photon 
direct transitions: The incoming
laser field was tuned to the half of the
excitonic frequency, so that the two-photon
transition was in exact resonance with
the triplet excitons. Such a method
allows to create a dense and coherent 
cloud of the triplet excitons. This
is practically a direct mean of creating
a condensate of the triplet excitons.
However,
fast collision induced 
decay of the triplet excitons may destroy
this coherence on the scale of few
nanoseconds. On one hand, 
this is the case if
the Auger decay is responsible for 
the triplet exciton depletion because
this channel is not sensitive to the 
orientation of the angular momenta of
the colliding pairs. On the other hand,
if the primary 
channel for the decay is
the mechanism \cite{KAV}, it should be
possible to use such a polarization
of the cloud that the created 
condensate of the triplet excitons
is stable on much longer time scale.

It is possible to employ general
symmetry considerations, and make 
a suggestion for the choice of 
the orientation of the incoming laser
fields in the geometry of the
experiment \cite{GOTO}. Indeed,
the two-photon process 
of creation of the triplet
exciton corresponds
to the interaction term
in the energy density 

\begin{eqnarray}
H_{le}=\sum_{a=\pm 1,0;i,j}
\psi^{\dagger}_{(a)}Q^{(a)}_{ij}E_iE_j +H.c.
\label{1}
\end{eqnarray}
\noindent
where $\psi_{(a)}$ stands for the 
triplet exciton Bose field which
has three projections $a=\pm 1$
and $a=0$ of the (internal)
angular momenta; $E_j$ denotes
three space components of the incoming
laser field $\sim E_j\exp (-i\omega t)$
which is taken in the
rotating wave approximation;
$ Q^{(a)}_{ij}$ are corresponding matrices
representing the point symmetry
(including spins) of $Cu_2O$ in such a way
that (\ref{1}) is invariant under
this symmetry. The interaction
term responsible for the decay \cite{KAV}
can be represented in the contact 
form (S-wave channel) as

\begin{eqnarray}
H_{op}=g_{op}\psi^{\dagger}\psi^{\dagger}
(\psi_{(+1)} \psi_{(-1)}+ \psi_{(-1)} 
\psi_{(+1)} +
\label{2} \\
+ \psi_{(0)} \psi_{(0)}) + H.c.
\nonumber
\end{eqnarray}
\noindent
where $\psi$ is the field 
of the singlet excitons, and
$g_{op}$ is the interaction constant
such that the rate estimated
in \cite{KAV} is $\sim g_{op}^2$.
It is worth noting that the term
in the brackets in (\ref{2})
is invariant under
the symmetry group (including spins)
of $Cu_2O$,
where the excitonic states are formed
on the total angular momenta states
of $Cu$ (see in \cite{CUO}). If the
triplet excitons are created in such
a manner that this invariant is zero,
the interconversion process will be 
 suppressed. 

The induced fields $\psi_{(a)}$ are
given from (\ref{1}) as
$ \psi_{(a)}\sim \sum_{i,j}
Q^{(a)}_{ij}E_iE_j $. If  
substituted into
(\ref{2}), this will result in the
term describing four-photon
production of the singlet
excitons which in general should be significant
as long as
the mechanism \cite{KAV}
is dominant, provided the 
density $|\psi_{(a)}|^2
\sim |\sum_{i,j}
Q^{(a)}_{ij}E_iE_j |^2$
 of the induced triplet excitons is
large enough. In fact, the
symmetry of $Q^{(a)}_{ij}$ is the
same as that of the tensors of the
direct quadrupole transitions for the
triplet excitons. Using this, it is possible
to find the energy density (\ref{2}) as  

\begin{eqnarray}
H_{op}\sim g_{op}\psi^{\dagger}\psi^{\dagger}
(E_x^2E_y^2 + E_x^2E_z^2+ E_y^2E_z^2) + H.c.
\label{3}
\end{eqnarray}
\noindent
where $E_x,\, E_y,\, E_z$ refer to the
components
of the laser field with respect to
the principal cubic axes of $Cu_2O$.
Accordingly,
the requirement 

\begin{eqnarray}
E_x^{-2}+E_y^{-2} + E_z^{-2}=0
\label{4}
\end{eqnarray}
\noindent
insures that the interconversion
process \cite{KAV} described
by (\ref{2}), (\ref{3}) is zero in the
dominant s-wave channel as long as no
thermalization of the created triplet
excitons occurs.

A solution of (\ref{4}) for the 
six components of $E_j=E'_j + iE''_j$,
where $ E'_j $ and $E''_j$ stand for
the real and imaginary parts of $E_j$,
respectively, can be represented as 
follows

\begin{eqnarray}
E_j={1\over \epsilon'_j +i\epsilon''_j}
\label{5}
\end{eqnarray}
\noindent
in terms of the two auxiliary
real vectors $\epsilon'_j,\, \epsilon''_j$
which are arbitrary except for the
conditions

\begin{eqnarray}
\sum_j \epsilon'^2_j=\sum_j\epsilon''^2_j, \quad 
\sum_j \epsilon'_j\epsilon''_j=0.
\label{6}
\end{eqnarray}
\noindent
The interpretation of these conditions is
straightforward: the complex vector $E_j$
represented by (\ref{5}) 
should be chosen in such a way that
the two auxiliary vectors $\epsilon'_j$
and $\epsilon''_j$ are equal in magnitude
to each other and are mutually orthogonal.
For the case of the incidence of the light 
along the direction (1,1,1), the solution
of (\ref{6}), (\ref{5}) gives 
that $\sum_j E'_j E''_j=0$ and
$\sum_j  E'^2_j=\sum_j  E''^2$.
Note that, given this, the
interconversion rate $k(T=0)=0$, as opposed
to the rate of the 
Auger decay which is not sensitive to
temperature and the mutual orientation
of the excitonic
angular momenta of the colliding pairs.

At finite temperatures $T\neq 0$,
the normal component - thermal triplet excitons -
is present in addition to the 
condensate. This component
should be characterized by zero net spin
polarization due to the interaction with phonons.
Thus, the interconversion process \cite{KAV}
will take place. Its rate $k(T)$ is proportional
to the normal density $n'$. Thus, it 
must be strongly temperature dependent. 
It is straightforward to find an estimate for $k(T)$ 
in the temperature range 
$\mu(0)<T \ll T_0$,
where $\mu(0)=4\pi an_0$ and $a,\, n_0$
stand
for the excitonic scattering length
and the exciton condensate (spin-polarized)
density, respectively; and $T_0$ denotes
the temperature of the excitonic Bose-Einstein
condensation. Indeed, in this range 
the normal component
behaves almost as an ideal gas. Thus, the
estimate follows from Eqs.(8-10) of Ref.
\cite{KAV} where the total density is replaced
by the density of the normal component
$n'=n_0(T/T_0)^{3/2}$ \cite{BEC}. Accordingly,
the ratio of the rate of the interconversion
$k(T)$ at $T\neq 0$ for the spin-polarized excitonic
condensate to the rate $1/\tau_{o,p}$ \cite{KAV}
estimated for the case of the non-polarized
cloud is

\begin{eqnarray}
k(T)\tau_{o,p}\approx {n^{\prime}\over n_0}=
\left({T\over T_0}\right)^{3/2} \ll 1
\end{eqnarray}
\noindent
for the temperatures under consideration.
At temperatures $T<\mu(0)$, further significant
reduction
of the rate should occur due to the interaction
between the triplet exciton condensate and the normal component.

\end{document}